# Few-layer hyperbolic multilayer for spontaneous emission enhancement

Ling Li,[1] Changjun Min,[1,*] Xiaocong Yuan[1,2]

[1]*Nanophotonics Research Centre, Shenzhen Key Laboratory of Micro-Scale Optical Information Technology & Institute of Microscale Optoelectronics, Shenzhen University, Shenzhen 518060, China*
*\*Corresponding author: cjmin@szu.edu.cn*
[2]*xcyuan@szu.edu.cn*

**Multilayer hyperbolic metamaterials consisting of alternating metal and dielectric layers have important applications in spontaneous emission enhancement. In contrast to the conventional choice of at least dozens of layers in multilayer structures to achieve tunable Purcell effect on quantum emitters, our numerical calculations reveal that multilayers with fewer layers and thinner layers would outperform in Purcell effect. These discoveries are attributed to the negative contributions by an increasing layer number to the imaginary part of the reflection coefficient, and the stronger coupling between surface plasmon polariton modes on a thinner metal layer. This work could provide fundamental insights and practical guide for optimizing the local density of optical states enhancement functionality of ultrathin and even two-dimensional photon sources.**

Purcell effect describes the amplification of a photon emitter's spontaneous emission rate by its local electromagnetic environment [1]. In pursuit of important implications of Purcell effect in a plethora of applications from micro- and nanoscale lasing, nanophotonics to quantum computation and quantum optics, many structures and materials have been proposed and realized [2-6]. These structures include resonant electric circuits, ultrasmall cavities, nano-antennas, and subwavelength metamaterials [7-11], among which multilayer based hyperbolic metamaterials have received a surge of research efforts for their open hyperbolic dispersions yielding an enormously large local density of optical states (LDOS) [11-17]. The attractive features of multilayer hyperbolic metamaterials include straightforward design and fabrication, and tunable LDOS enhancement realized by adjusting the epsilon-near-zero (ENZ) spectral position [18, 19]. The appearance of ENZ would induce the insulator-to-metallic transition of the whole multilayer metamaterials accompanied by an LDOS enhancement peak. The metal-to-dielectric ratio controls the ENZ spectral position and thus the LDOS enhancement peak of metal-dielectric multilayers. To design such multilayer metamaterials, care must be taken regarding the metal and dielectric layer thickness and the numbers of alternating layers to make the effective-medium theory (EMT) an accurate design tool [20, 21].

Hyperbolic multilayers reported so far consist of at least dozens of alternating metal and dielectric layers to enhance spontaneous emission of photon emitters [12, 13, 15, 18, 19]. However, the large number of layers has not been justified in current implementations of hyperbolic multilayers for enhancing spontaneous emission rates. Investigation on the effects of layer number and thickness of multilayer metamaterials on the LDOS enhancement is still lacking. It is not clear whether and how would these layers contribute at increasing distances from the photon emitters. Settling these concerns would open up the potential for optimizing the Purcell effect via adjusting the layer number and thickness.

In this work, we investigate the effects of layer number and thickness on the Purcell effect of a multilayer composed of alternate silver (Ag) and silicon (Si) layers on a dipolar emitter on top of it. Surprisingly, we found that the multilayer structures with a few layers of small thickness outperform those with many layers of large thickness in terms of the Purcell effect. The theoretical analysis attributes the influence of layer number on Purcell effect to the negative contributions of those layers following the topmost one to the imaginary part of the complex reflection coefficient, and the layer thickness effect to the dispersion of coupled surface plasmon polariton (SPP) modes at the two metal-dielectric interfaces of a thin metal layer. This work would establish an optimization procedure for simultaneously maximizing the Purcell effect of planar few-layer metamaterials and minimizing fabrication demand. Moreover, it can also provide insights into exploring ultrathin structures such as metasurfaces and two-dimensional materials for novel photon sources.

The periodically stratified multilayer studied in this work is illustrated in the left panel of Fig. 1(a). Its unit-cell starts with an Ag layer followed by a Si layer with thicknesses denoted by $d_m$ and $d_d$, respectively. The aforementioned layer number effect is studied by calculating the Purcell factors (PF) experienced by a dipole polarized perpendicularly to and located at a distance of $h = 5$ nm from the topmost Ag layer of a series of multilayers with an increasing number of layers. The layer number is increased by alternatively adding a single Ag or Si layer along the dashed arrow. The multilayer structure is sandwiched by the air below and a polymethyl methacrylate (PMMA) superstrate where the dipole resides. In this scenario, the PF for the dipole is [22]

$$Re\left[\int_0^\infty \frac{3}{2\sqrt{\varepsilon_1 k_0^2 - k_x^2}} \left(\frac{k_x}{k_0\sqrt{\varepsilon_1}}\right)^3 (1 + r^p e^{2i\sqrt{\varepsilon_1 k_0^2 - k_x^2}h})dk_x\right], \quad (1)$$

where the integrand is the total LDOS enhancement from the multilayer using Fermi's golden rule as an analog [11], $\varepsilon_1$ is the relative permittivity for PMMA, $k_0$, $k_x$, and $\sqrt{\varepsilon_1 k_0^2 - k_x^2}$ is the vacuum optical wavevector, wavevector component parallel and perpendicular to the multilayer top surface, respectively, and $r^p$ is the multilayer's reflection coefficient for p-polarized incident waves. Using dielectric permittivity of all materials [23-25] and $r^p$ obtained by the rigorous transfer matrix method (TMM) [26], Eq. 1 was employed in MATLAB (MathWorks) to calculate PFs. The right panel of Fig. 1(a) shows PF spectra, using TMM, from two typical multilayer examples which consist of 2 layers (one unit-cell) with $d_m = d_d = 1$ nm and of 60 layers (30 unit-cells) with $d_m = d_d = 40$ nm, denoted by the black dashed and violet solid lines, respectively. The numerical calculations based on TMM are verified by simulations using the three-dimensional finite-difference time-domain method (Lumerical FDTD Solutions) represented by red circles. We also calculated the PFs for the two multilayers using the EMT to obtain the reflection coefficient for the equivalent layers, which are denoted by the black and violet crosses, respectively. For the one unit-cell with the 1 nm layers, the EMT result agrees very well with that obtained using the TMM, while that of the 30 period of unit-cells with the layer thickness of 40 nm does not. This deviation between the EMT and the accurate TMM is supported by studies on the limitations of EMT in describing multilayer optics [20, 21, 27].

Figs. 1(b) and 1(c) depict the peak values and peak positions of the PFs experienced by the electric dipole shown in Fig.1(a). In Fig. 1(b), for fixed layer thicknesses, multilayers with only one unit-cell or a few layers usually offer the most significant PF peak values compared to those with more layers. And the multilayer with the thinnest layer has the largest PF peak values in the hyperbolic region among all multilayers. For instance, one period of the unit-cell with a layer thickness of 1 nm offers the largest PF value of 3932. In contrast, for the multilayer with a layer thickness of 40 nm, the largest PF is only 16. Counterintuitively, extra layers underneath contribute negatively to the overall LDOS enhancement effect. Moreover, this negative PF contribution is more profound for the thinner layer. For the layer thickness of 1 and 40 nm, the PF peak value of the multilayer with 60 layers (30 unit-cells) is about 5.8 and 1.4 times smaller than that of the Ag-Si and Ag-Si-Ag-Si multilayers, respectively. The PF peak positions illustrated in Fig. 1(c) fluctuate. Since the PF peaks around ENZ wavelength, for multilayers with more than just a few layers, these fluctuations can be understood from the EMT, from which the ENZ wavelength can be derived by locating where effective permittivity components change signs [20]. As layer number increases, the Ag filling ratio changes and shifts the ENZ wavelength for every multilayer. For few-layer structures, the ENZ spectral position strongly depends on the layer numbers and waveguide modes [21, 28]. Thus the fluctuation is more substantial in contrast to the thicker layer multilayers. Towards large layer numbers (as shown in the inset of Fig. 1(c)), the degree of this fluctuation for multilayers with thinner layers decreases but still shows a level stronger than that of the multilayers with thicker layers. The behavior of PF presented in Fig.1 also exists for a horizontally polarized dipole, whose results are not shown here.

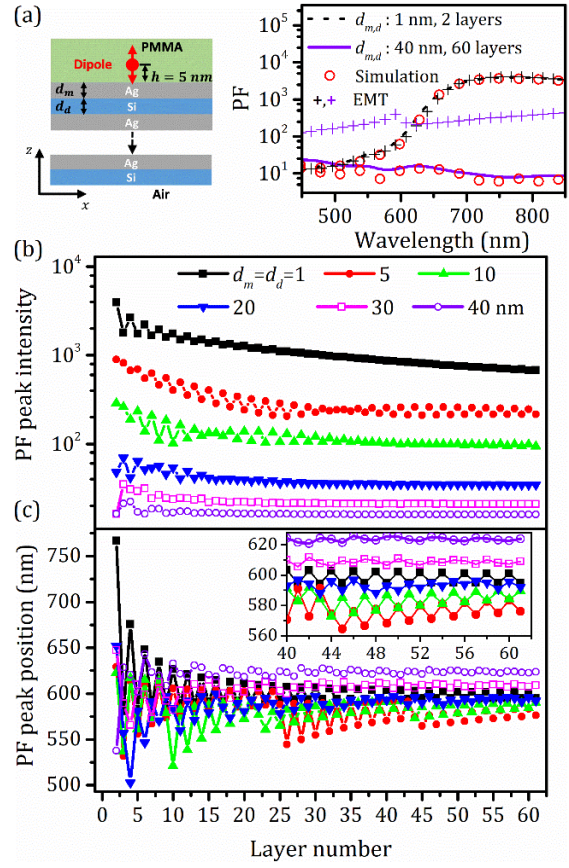

Fig. 1. (a) Multilayer sandwiched between PMMA and air consists of Ag (thickness $d_m$) and Si ($d_d$) layers. The dipole is 5 nm above the multilayer and polarized along z-direction (left). PFs of one multilayer with 2 layers of Ag and Si with $d_m = d_d = 1$nm and another with 60 layers with $d_m = d_d = 40$nm (right). Lines: calculations using TMM, red circles: FDTD simulation, crosses: calculations using EMT. (b) PF peak intensity and (c) peak position of multilayers with different layer thicknesses and layer numbers (starting from 2). Layer thicknesses of 1, 5, 10, 20, 30, and 40 nm are denoted by different line styles, as shown in (b).

To understand the effect of layer number, we should notice that the dominating contribution to the PF comes from the high-$k$ hyperbolic modes and that at very large $k_x$ the value of Eq. 1 is directly proportional to the imaginary part of the reflection coefficient [18, 22]. The p-wave reflection coefficients for multilayers with $N$ and $N$-1 periods of unit-cell are related by the propagation term associated with the unit-cell [26], i.e., $r_N^p = r_{N-1}^p e^{-2i\beta}$, where $\beta$ is the complex propagation phase along the unit-cell [26]. Following the wave form convention in [26], we can define $r_{N-1}^p = a_{N-1} + ib_{N-1}$ and $\beta = K - iJ$ where $a_{N-1}, b_{N-1}, K$ are real-valued, and $J > 0$. After some simple algebra and notice that the LDOS enhancement peaks near the Bloch band edge where $\cos k \approx 1 \gg \sin k$, we have the imaginary part of the reflection coefficient

$$Im(r_N^p) \approx b_{N-1}e^{-2J} < Im(r_{N-1}^p). \quad (2)$$

Eq. 2 is the key result of this work; it reveals that the source of the layer number effect observed in Fig. 1 is the decaying nature of the evanescent SPP mode across the layers in the high-$k$ hyperbolic region. To demonstrate the layer number effect explicitly, in Fig. 2, we plot the $Im(r_N^p)$ as a function of period number $N$ using the accurate TMM with 10 nm thickness for both the Ag/Si layer at the wavelength of 600 nm. The overall decaying trend depicted in Fig. 2 agrees well with that of the PF with even layer numbers illustrated in Fig. 1(b). The trend of the PF with odd layer numbers can also be examined and explained similarly, which are not shown here.

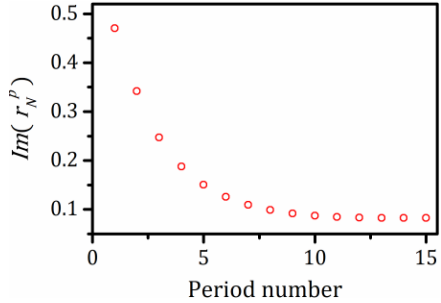

Fig. 2. The imaginary part of the reflection coefficient of multilayers of 10 nm layers as a function of the period (unit-cell) number $N$.

The dependence of LDOS enhancement on layers' thickness can be understood from the mechanisms underlying the coupled SPP modes of a thin metallic layer [29, 30]. To illustrate this point, Figs. 3(a) and 3(b) show the wavevector-layer thickness dispersions of coupled SPP modes on an Ag layer at the ENZ wavelength of 600 nm calculated using EMT for multilayers with equal Ag/Si layer thicknesses. The Ag layer is sandwiched by a PMMA superstrate and a Si substrate. The long-range SPP (LRSPP) and short-range SPP (SRSPP) modes are denoted by hollow green circles and solid red circles. They separate further from each other for the Ag layer thinner than 20 nm. Both the real and imaginary wavevector parts of the SRSPP mode increase dramatically towards thinner layers, leading to more confined SPP fields and stronger Purcell effect of plasmonic systems [29, 30]. This layer thickness dependence of coupled SPP modes also dictates their contributions to the Ag layer's LDOS enhancements, shown in Figs. 3(c) and 3(d) where the energy-wavevector dispersions of the coupled SPP modes marked by circles are overlapped with the LDOS enhancement spectra of a 20 and 5 nm thick Ag layer, respectively. The SRSPP (LRSPP) modes are solid red circles (hollow green circles) and lie at the lower (higher) energy region approaching from bellow (above) toward $\omega_{s1} = 0.23\omega_p$ ($\omega_{s2} = 0.38\omega_p$). $\omega_{s1}$ and $\omega_{s2}$ are the surface plasmon resonance frequencies at the Ag-Si and Ag-PMMA interfaces, where $\omega_p$ is the bulk plasmon resonance frequency of Ag (see the two dash-dot lines in Fig. 3(d)). For the 20 nm Ag layer case shown in Fig. 3(c), the LRSPP mode does not align well with the LDOS spectrum at $\omega_{s2}$, the major LDOS enhancement contribution around $\omega_{s1}$ and $\omega_{s2}$ comes from the SRSPP mode and the intrinsic surface modes (not shown here), respectively. In contrast, both LRSPP and SRSPP modes coincide with the LDOS spectrum very well, as shown in Fig. 3(d), indicating the SPP coupling strength in 5 nm Ag layer is much stronger. For a thinner Ag layer, the stronger SPP coupling strength increases the wavevector range for both LRSPP and SRSPP modes. However, only the spectral intensity of LDOS originated from the SRSPP mode is enhanced. The LRSPP mode at $\omega_{s2}$ not only weakens the intensity of the LDOS but also lifts the LDOS mode curve, making it less flat, which would reduce the contribution to the integrated PF. The above analysis agrees with previous studies revealing that the SRSPP (LRSPP) contribution to the LDOS increases (decreases) as the metal layer gets thinner [29, 30].

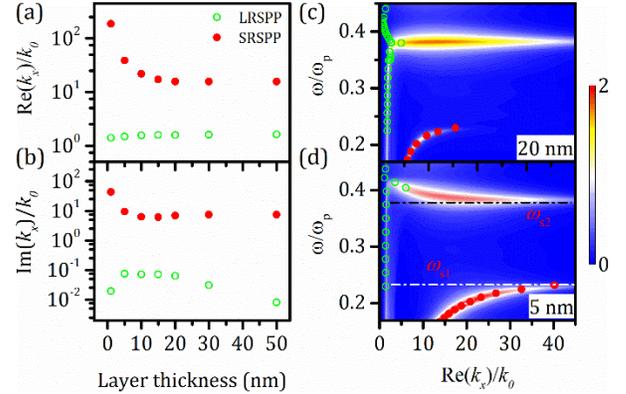

Fig. 3. (a) The real and (b) imaginary parts of the normalized propagation wavevectors of LRSPP (hollow green circles) and SRSPP (solid red circles) modes at the wavelength of 600 nm on an Ag layer sandwiched by a PMMA superstrate and a Si substrate as a function of Ag layer thickness. (c) and (d) depict the energy-wavevector dispersions of the coupled SPP modes (in circles) overlapped with the LDOS enhancement spectra of a 20 and 5 nm thick Ag layer, respectively.

In multilayer structures, the addition of dielectric layers would alter the LDOS spectra from one metal layer shown in Fig. 3. However, the above analysis also applies. Figs. 4(a) and 4(b) compare the PF spectra which peak around the SRSPP and LRSPP modes, respectively, of the one Ag-Si unit-cell with 5, 10, and 20 nm layer thicknesses. The PF around the wavelength of 600 nm ($\sim\omega_{s1}$) and 350 nm ($\sim\omega_{s2}$) increases and decreases, respectively, as the layers' thickness decreases. The LDOS enhancement for the 5 and 20 nm layer thickness cases are illustrated in Figs. 4(c) and 4(d). By comparing the LDOS enhancement spectra, we can observe that the SRSPP nature is prevailing in the lower energy ($\omega_{s1}$) region with mode intensity inversely proportional to the layer thickness. On the other hand, the modes in the higher energy region around $\omega_{s2}$ show a feature resembling the LRSPP when the layer gets thinner. PF behavior in Figs. 4(a) and 4(b) can be clarified by this layer thickness dependence of coupled SPP modes and the fact that an SRSPP mode surpasses the LRSPP mode at contributing to a PF at a specific spectral position [29, 30].

Furthermore, we can control the PF peak value and position by adjusting the metal filling ratio. Fig. 5 shows the PFs of one unit-cell with different Si thickness while keeping the Ag layer thickness equal to 10 nm. We observe that the PF can be tuned across a wide wavelength range using even a unit-cell structure. Thinner Si layers give stronger PFs at shorter wavelengths, while thicker Si layers exhibit lower PF peak values, but broader PF enhancement ranges reaching into longer wavelength regions. In principle, these observations can also be explained by those analyses of coupled SPP applied to results in Figs. 3 and 4. This broadband tunable LDOS enhancement capability highlighted as a crucial feature of conventional hyperbolic multilayers with many layer numbers, achieved here by designing few-layer systems, is appealing to many applications with less demanding fabrications.

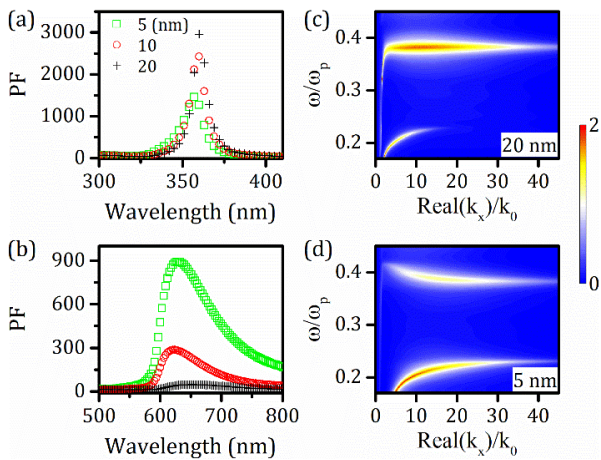

Fig. 4. PFs of the one Ag-Si unit-cell with 5 nm (hollow green squares), 10 nm (hollow red circles), and 20 nm (black crosses) layer thickness at short wavelength (a) and long wavelength (b). (c), (d) LDOS enhancement spectra of one Ag-Si unit-cell with the two different layer thicknesses.

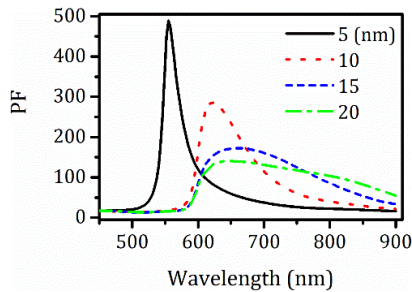

Fig. 5. PFs of one unit-cell of Ag and Si layers with Ag layer thickness fixed to 10 nm and Si layer thickness varying from 5 to 10, 15, and 20 nm.

In summary, we discovered and explicitly elucidated the counter-intuitive phenomena that the optimal layer number and layer thickness to achieve strong PF for multilayer based metamaterials is limited to a few and extremely small, respectively. The extra layers beyond this optimal layer number would induce negative PF contributions with an oscillatory and decaying magnitude as the layer number increases. The concept of optimal layer number and thickness in planar metamaterial systems, as unexplored in conventional multilayer metamaterials, would bring invaluable insights and practical guidance to developing thin and compact or even two-dimensional materials with tunable enhancement on LDOS [31]. Future explorations include considering the effects of quantum confinement, nonlocal dielectric response, and finite emitter size on the LDOS enhancement and other light-matter interactions from the few-layer multilayers with ultrathin layers.

**Funding.** National Natural Science Foundation of China (NSFC) (91750205, U1701661, 61935013, 61805165); Leading Talents of Guangdong Province Program (00201505); Natural Science Foundation of Guangdong Province (2016A030312010); Science and Technology Innovation Commission of Shenzhen (JCYJ20180507182035270, KQTD2017033011044403, ZDSYS201703031605029).

**Disclosures**. The authors declare no conflicts of interest.